\documentclass[prb,showpacs,twocolumn]{revtex4}

\usepackage{epsfig}

\usepackage{graphics}

\usepackage{graphicx}

\usepackage{dcolumn}

\usepackage{bm}

\newcommand{\sba}{\begin{subeqnarray}}

\newcommand{\sea}{\end{subeqnarray}}

\def\cm-1{cm$^{-1}$}

\begin{document}

\title{Diagrammatic theory for Periodic Anderson Model.\\Stationary property of the thermodynamic potential}
\author{V.\ A.\ Moskalenko$^{1,2}$}
\email{moskalen@theor.jinr.ru}
\author{L.\ A.\ Dohotaru$^{3}$}
\author{R.\ Citro$^{4}$}
\affiliation{$^{1}$Institute of Applied Physics, Moldova Academy
of Sciences, Chisinau 2028, Moldova} \affiliation{$^{2}$BLTP,
Joint Institute for Nuclear Research, 141980 Dubna, Russia}
\affiliation{$^{3}$Technical University, Chisinau 2004, Moldova}
\affiliation{$^{4}$Dipartimento di Fisica E.\ R.\ Caianiello,
Universit\'{a} degli Studi di Salerno and CNISM, Unit\'{a} di
ricerca di Salerno, Via S. Allende, 84081 Baronissi (SA), Italy}
\date{\today}

\begin{abstract}


%

Diagrammatic theory for Periodic Anderson Model has been
developed, supposing  the Coulomb repulsion of $f-$ localized
electrons as a main parameter of the theory. $f-$ electrons are
strongly correlated and $c-$ conduction electrons are
uncorrelated. Correlation function for $f-$ and mass operator for
$c-$ electrons are determined. The Dyson equation for $c-$ and
Dyson-type equation for $f-$ electrons are formulated for their
propagators. The skeleton diagrams are defined for correlation
function and thermodynamic functional. The stationary property of
renormalized thermodynamic potential about the variation of the
mass operator is established. The result is appropriate as for
normal and as for superconducting state of the system.
\end{abstract}

 \maketitle

\section{Introduction}

The study of the systems with strongly correlated electrons has
become in the last time one of the central problem of condensed
matter physics. One of the most important models of strongly
correlated electrons is  periodic Anderson model (PAM)$^{[1]}$.
This model is used to describe the physics of mixed  valence
systems, heavy fermion compounds, high-temperature
superconductivity as well as other phenomena in which the strong
Coulomb repulsion of the localized electrons is present. This
model describes the intermetallic compounds which contain magnetic
moments of rare earth or actinide ions  included in the host
metal. This ions have a partially filled $f-$ shell and can be
considered as scattering centers for conduction electrons of the
host metal. Because of the strong Coulomb repulsion of the
electrons with opposite spins located at the same site of lattice
the magnetic ion electrons are strongly correlated. There is also
the hybridization of states between the uncorrelated conduction
electrons and localized correlated ones when both of them are
present on the same lattice site. Magnetic properties of the
impurities ions affect in a different manner the properties of the
host matrix and of the system as a whole. For different regime of
physical parameters, determined by Coulomb local interaction,
hybridization of the wave functions and exchange interaction, it
is possible to obtain different classes of the system phases.

There are already an enormous number of approximate methods and
approaches devoted to PAM, as perturbation expansions, static and
dynamic mean field theories, variational and numerical approaches,
large $N$ expansion , slave boson methods, non crossing
approximations (NCA), Bogoliubov inequality method and others.
Also some exact results are known, obtained in special with the
Bethe ansatz, renormalization group methods and Bogoliubov
inequality method. We will not enlarge upon the most essential
stages in the development of this model because exists a number of
consistent reviews $^{[2-7]}$ and books $^{[8,9]}$ on this field
and we shall use the references to previous our papers.


\section{Model Hamiltonian}


We consider the simplest form of PAM with a spin degeneration of
the level of localized $f-$ electrons, a simple energy band of
conducting $c-$ electrons, Coulomb one-site repulsion $U$ of
correlated $f-$ electrons with opposite spins and one-site
hybridization between both group of electrons of this system. The
hamiltonian of the system reads:
%
\begin{eqnarray}
H&=&H_{c}^{0}+H_{f}^{0}+H_{int},  \nonumber \\
H_{c}^{0}&=&\sum\limits_{\mathbf{k}\sigma }\epsilon (\mathbf{k})\
C_{\mathbf{k}\sigma }^{+}C_{\mathbf{k}\sigma },  \nonumber \\
H_{f}^{0}&=&\epsilon _{f}\sum\limits_{i\sigma }n_{i\sigma }
+U\sum\limits_{i}n_{i\uparrow }n_{i\downarrow }, \\
H_{int}&=&V\sum\limits_{i\sigma }\left(C_{i\sigma }^{+}f_{i\sigma
} + f_{i\sigma }^{+}C_{i\sigma }\right) ,  \nonumber \label{1}
%
\end{eqnarray}
where
%
\begin{eqnarray}
\epsilon
(\mathbf{k})&=&\overline{\epsilon}(\mathbf{k})-\mu,\nonumber
\epsilon _{f}=\overline{\epsilon} _{f}-\mu, \nonumber\\
n_{i\sigma}&=& f_{i\sigma }^{+}f_{i\sigma}, \\C_{i\sigma
}&=&\frac{1}{\sqrt{N}}\sum\limits_{\mathbf{k}}
\exp{(-i\mathbf{k}\mathbf{R_{i}})}C_{\mathbf{k}\sigma}.\nonumber\label{2}
%
\end{eqnarray}

Here $V$ is the hybridization amplitude assumed constant. We have
indicated with $C_{i\sigma }^{+}(f_{i\sigma }^{+})$ the creation
operator for an uncorrelated (correlated) electron with spin
$\sigma$ and $i$ lattice site, $n_{i\sigma}$ is the number
operator for $f-$ electrons, $\epsilon (\mathbf{k})$ is the band
energy with momentum $\mathbf{k}$ of conductivity electrons spread
on the entire width $W$ of the band.  $\epsilon _{f}$ is the
energy of localized electrons. Both these energies are evaluated
with respect to the chemical potential $\mu$.

The approach proposed in this paper generalizes the diagrammatic
theory of normal and superconducting phases of strongly correlated
systems proposed in previous papers $^{[10-19]}$.

The strong on-site repulsion $U$ between $f-$ electrons of
opposite spins is the main term in the Hamiltonian.

As the conduction electrons can belong not only to the $s-$ but
also to the $d-$ atomic shell, their Coulomb repulsion can also be
important. In this case the extended PAM must be used $^{[20]}$.
For simplicity the correlations of $c-$ electrons are not
considered and one subsystem is of $c-$ uncorrelated and the
second of $f-$ correlated electrons. Because of strong
localization of the $f-$ electrons they cannot hope from one
lattice site to another and their delocalization is due only to
the hybridization of the $f-$ and $c-$ states with matrix element
$V$. It is obvious that at $V\neq 0$ in the given model with two
subsystems superconductivity arises simultaneously in both
subsystems.

In the present paper we develop the thermodynamic perturbation
theory for the system in the superconducting state with
Hamiltonian (1) under the assumption that the term responsible for
hybridization of $c-$ and $f-$ electrons is a perturbation.

The Hamiltonian $H_{c}^{0}$ of the uncorrelated $c-$ electrons is
diagonal in band representation, where as the Hamiltonian
$H_{f}^{0}$ is diagonalized by using Hubbard transfer operators
$^{[21]}$. Therefore in the zeroth-order of the perturbation
theory the statistical operator of grand canonical ensemble of the
system is factorized in the momentum representation for $c-$ and
in local representation for $f-$ electrons:
%
\begin{eqnarray}
%
\exp{[-\beta(H_{c}^{0}+H_{f}^{0})]}&=&\prod\limits_{\mathbf{k}\sigma}\exp{(-\beta
H_{c\sigma}^{0}(\mathbf{k}))}\times \nonumber\\
\times\prod\limits_{i}\exp{(-\beta H_{f}^{0}(i))},\nonumber\\
H_{c\sigma}^{0}(\mathbf{k})&=&\epsilon (\mathbf{k})\
C_{\mathbf{k}\sigma }^{+}C_{\mathbf{k}\sigma },  \\
H_{f}^{0}(i)&=&\epsilon _{f}\sum\limits_{\sigma }n_{i\sigma }
+Un_{i\uparrow }n_{i\downarrow }\nonumber, \label{3}
%
\end{eqnarray}
%

We use the series expansion for evolution operator:
%
\begin{eqnarray}
%
U(\beta ) & = & T\exp (-\int\limits_{0}^{\beta }H_{int}(\tau
)d\tau ).\label{4}
%
\end{eqnarray}
%
in the interaction representation for electron operators
($a=c,f$):
%
\begin{eqnarray}
%
a(x)= e^{\tau H^{0}}a(\mathbf{x})e^{-\tau H^{0}},\overline{a}(x)=
e^{\tau H^{0}}a^{+}(\mathbf{x})e^{-\tau H^{0}}.\label{5}
%
\end{eqnarray}
%
Here by $x$ means ($\mathbf{x},\sigma,\tau$).

We shall denote by $\left\langle TAB...\right\rangle _{0}$ the
thermodynamic average with zeroth-order statistical operator (3)
of the chronological product of electron operators $(AB...)$. Such
averages are calculated independently for $c-$ and $f-$ operators
with using for $c-$ electrons the Wick Theorem of weak quantum
field theory and by using for $f-$ electrons the Generalized Wick
Theorem (GWT) proposed by us in papers $^{[10-13]}$ for strongly
correlated electron systems.

In the superconducting state, unlike the normal one, nontrivial
statistical averages of operator products with even total number
but inequal number of creation and annihilation electron operators
are possible. They realize the Bogoliubov quasi-averages $^{[22]}$
or Gor'kov $^{[23]}$ anomalous Green's functions. To unify the
calculation of statistical averages for normal and superconducting
phases it is useful to assign  an additional quantum number
$\alpha$, called by us charge number $^{[15]}$, with the values
$\pm 1$, which can be add to electron operators according the rule
($a=c,f$):

\begin{eqnarray}
a^{\alpha}(\mathbf{x})=\left\{\begin{array}{c} a(\mathbf{x}),\quad \alpha=1; \\
a^{+}(\mathbf{x}),\quad \alpha=-1. \end{array} \right.
\end{eqnarray}

In this representation the interaction operator $H_{int}$ becomes:
%
\begin{eqnarray}
%
H_{int}&=&V\sum\limits_{i\sigma \alpha}\alpha f_{i\sigma
}^{-\alpha} C_{i\sigma }^{\alpha} . \label{7}
%
\end{eqnarray}
%

Obviously, introducing a new quantum charge number leads to
additional summation over it values in all diagram lines and to an
additional factor $\alpha$ in the vertices of diagrams.

Now, after such introducing, it is irrelevant whether one deals
with creation or annihilation operators. First of all we shall
enumerate the main results of diagrammatic theory obtained in the
previous paper $^{[15]}$ necessary to our proving of stationary
theorem. Such theorem for uncorrelated many-electron systems in
normal state has been proved by Luttinger and Word $^{[24]}$.

\section{Perturbative Treatment [15]}


We use the definition of the one-particle Matsubara Green's
functions for $c-$ and $f-$ electrons
%
\begin{eqnarray}
%
G_{\alpha \alpha^{\prime}}^{c}(x|x^{\prime})&=&-\ \left\langle
Tc^{\alpha}(x)c^{-\alpha {\prime}}(x^{\prime})U(\beta
)\right\rangle _{0}^{c}\nonumber,\\
&&\\
G_{\alpha \alpha^{\prime}}^{f}(x|x^{\prime})&=&-\ \left\langle
Tf^{\alpha}(x)f^{-\alpha {\prime}}(x^{\prime})U(\beta
)\right\rangle _{0}^{c},\nonumber \label{8}
%
\end{eqnarray}
%
where index $c$ for $\left\langle ...\right\rangle _{0}^{c}$ means
the connected  of the diagrams which are taken into account in the
right-hand part of definition (8).

The following condition is fulfilled
%
\begin{eqnarray}
%
G_{\alpha
\alpha^{\prime}}^{a}(x|x^{\prime})&=&-G_{-\alpha^{\prime},
-\alpha}^{a}(x^{\prime}|x),a=(c,f).\label{9}
%
\end{eqnarray}
%

Between this new definition and traditional one  $^{[23]}$ there
is a relation
%
\begin{eqnarray}
%
G_{1,1}^{a}(x|x^{\prime})&=&G^{a}(x|x^{\prime})\nonumber,\\
G_{1,-1}^{a}(x|x^{\prime})&=&F^{a}(x|x^{\prime}),\nonumber\,\\
G_{-1,1}^{a}(x|x^{\prime})&=&\overline{F}^{a}(x|x^{\prime})\nonumber,\\
G_{-1,-1}^{a}(x|x^{\prime})&=&-G^{a}(x^{\prime}|x).\label{10}
%
\end{eqnarray}
%
In the presence of strong correlations of $f-$ electrons the (GWT)
contains additional terms namely the irreducible one-site
many-particle Green's functions or Kubo cumulants of the form ($x=
\mathbf{x},\sigma,\tau$):
%
\begin{eqnarray}
%
G_{n}^{(0)ir}[\alpha_{1},x_{1};...;\alpha_{2n},x_{2n}]=\
\left\langle
Tf_{x_{1}}^{\alpha_{1}}...f_{x_{2n}}^{\alpha_{2n}}\right\rangle
_{0}^{ir}=\\
\nonumber=\delta_{\mathbf{x}_{1}\mathbf{x}_{2}}...\delta_{\mathbf{x}_{1}\mathbf{x}_{2n}}\
\left\langle
Tf_{\sigma_{1}}^{\alpha_{1}}(\tau_{1})...f_{\sigma_{2n}}^{\alpha_{2n}}(\tau_{2n}\right\rangle
_{0}^{ir}, \label{11}
%
\end{eqnarray}
%
where  in simplest two-particle case we have the following
definition of the irreducible function
%
\begin{eqnarray}
%
\left\langle
Tf_{\sigma_{1}}^{\alpha_{1}}(\tau_{1})f_{\sigma_{2}}^{\alpha_{2}}(\tau_{2})f_{\sigma_{3}}^{\alpha_{3}}(\tau_{3})
f_{\sigma_{4}}^{\alpha_{4}}(\tau_{4})\right\rangle
_{0}^{ir}&=&\\\nonumber =\left\langle
Tf_{\sigma_{1}}^{\alpha_{1}}(\tau_{1})f_{\sigma_{2}}^{\alpha_{2}}(\tau_{2})f_{\sigma_{3}}^{\alpha_{3}}(\tau_{3})
f_{\sigma_{4}}^{\alpha_{4}}(\tau_{4})\right\rangle
_{0}&-&\\\nonumber-[\left\langle
Tf_{\sigma_{1}}^{\alpha_{1}}(\tau_{1})f_{\sigma_{2}}^{\alpha_{2}}(\tau_{2})\right\rangle
_{0}\left\langle Tf_{\sigma_{3}}^{\alpha_{3}}(\tau_{3})
f_{\sigma_{4}}^{\alpha_{4}}(\tau_{4})\right\rangle
_{0}&+&\\\nonumber+\left\langle
Tf_{\sigma_{1}}^{\alpha_{1}}(\tau_{1})f_{\sigma_{4}}^{\alpha_{4}}(\tau_{4})\right\rangle
_{0}\left\langle Tf_{\sigma_{2}}^{\alpha_{2}}(\tau_{2})
f_{\sigma_{3}}^{\alpha_{3}}(\tau_{3})\right\rangle
_{0}&-&\\\nonumber-\left\langle
Tf_{\sigma_{1}}^{\alpha_{1}}(\tau_{1})f_{\sigma_{3}}^{\alpha_{3}}(\tau_{3})\right\rangle
_{0}\left\langle Tf_{\sigma_{2}}^{\alpha_{2}}(\tau_{2})
f_{\sigma_{4}}^{\alpha_{4}}(\tau_{4})\right\rangle
_{0}].\label{12}
%
\end{eqnarray}
%
For $n\geq3$ the irreducible Green's function contains in its
right-hand part besides the products of one-particle
propagators also their products with irreducible functions of
smaller number of particles. There are present also the product
of irreducible functions $G_{n_{1}}^{(0)ir}$ and
$G_{n_{2}}^{(0)ir}$ with condition $n_{1}+n_{2}=n$ and so on
$^{[10-13]}$.

As a result of applying  these theorem we obtain for the
renormalized conduction electron propagator the contributions
depicted on the Fig. 1
%
\begin{figure*}[t]
%
\centering
\includegraphics[width=0.85\textwidth,clip]{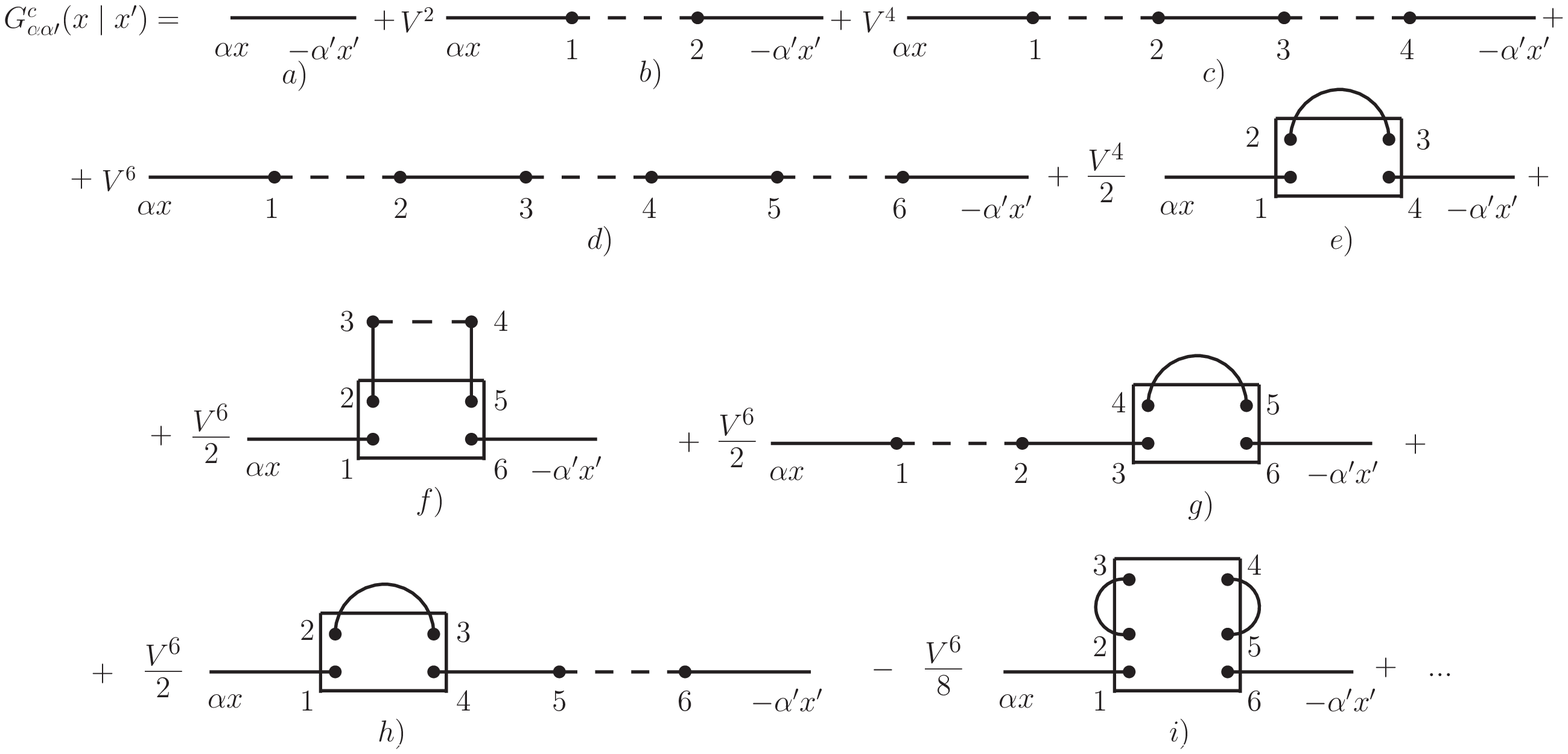}
\vspace{-0mm}
%
\caption{ The first six orders of perturbation theory for
conduction electron propagator. The solid and dashed thin lines
depict zero order propagators for $c-$ and $f-$ electrons
correspondingly. The rectangles depict the irreducible Green's
functions. The points of diagram are the vertices with $\alpha$
and $V$ contributions. }\label{fig-1} \vspace{-5mm}
\end{figure*}
%
The contributions of the diagrams Fig. 1 b) and e) are the
following
%
\begin{eqnarray}
%
V^{2}\sum\limits_{\alpha_{1}\alpha_{2}}\sum\limits_{12}\alpha_{1}\alpha_{2}G_{\alpha
\alpha_{1}}^{c(0)}(x|1)G_{\alpha_{1}
\alpha_{2}}^{f(0)}(1|2)G_{\alpha_{2}
\alpha^{\prime}}^{c(0)}(2|x^{\prime}),\nonumber\\
\frac{V^{4}}{2}\sum\limits_{\alpha_{1}...\alpha_{4}}\sum\limits_{1...4}\alpha_{1}\alpha_{2}\alpha_{3}\alpha_{4}
G_{\alpha\alpha_{1}}^{c(0)}(x|1)G_{\alpha_{2}
\alpha_{3}}^{c(0)}(2|3)\times\nonumber\\\times G_{\alpha_{4}
\alpha^{\prime}}^{c(0)}(4|x^{\prime})\left\langle
Tf_{1}^{\alpha_{1}}f_{2}^{-\alpha_{2}}f_{3}^{\alpha_{3}}f_{4}^{-\alpha_{4}}\right\rangle
_{0}^{ir}\nonumber\label{13}
%
\end{eqnarray}
%
correspondingly. It demonstrates the dependence of the diagrams
from the charge quantum number $\alpha$.

The contributions of perturbation theory for $f-$ electron
propagator are depicted on the Fig. 2
%
\begin{figure*}[t]
%
\centering
\includegraphics[width=0.85\textwidth,clip]{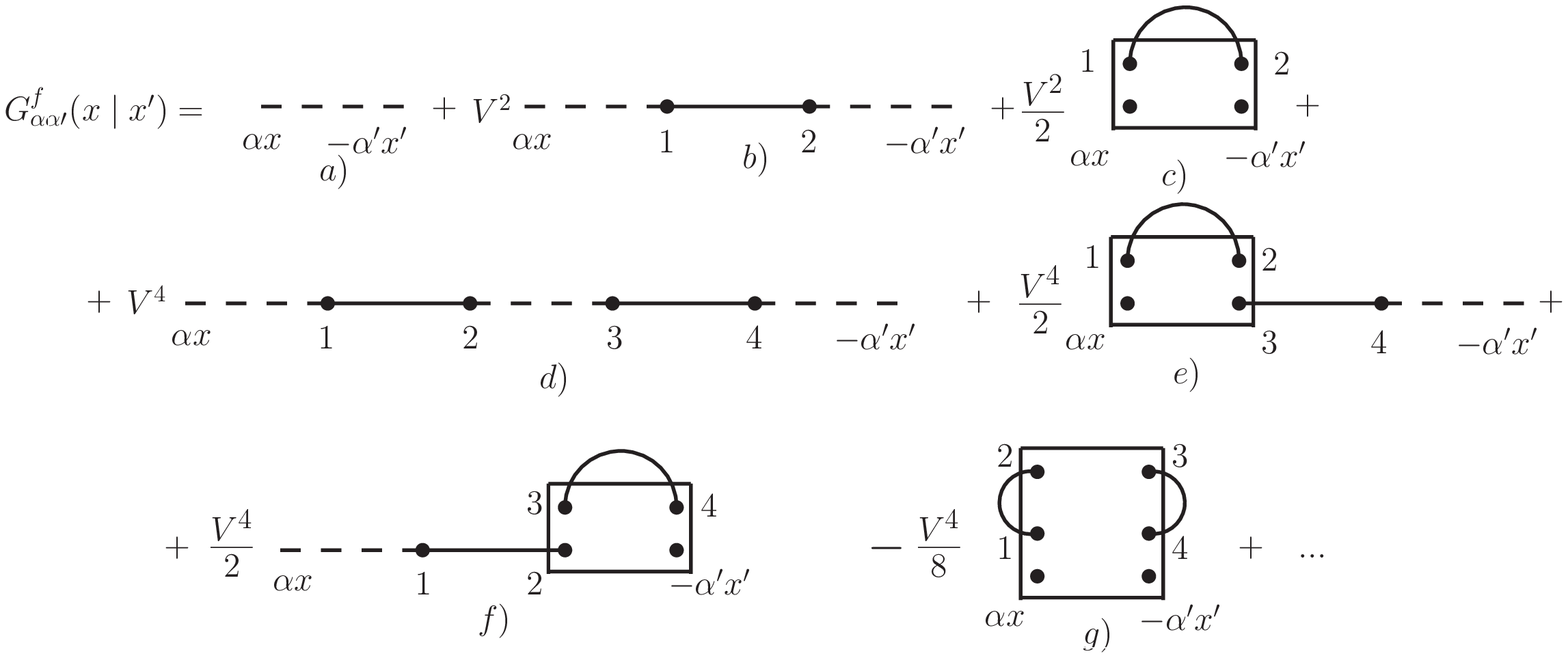}
\caption{ The contributions of the first four orders of
perturbation theory for the $f-$ electron propagator.
}\label{fig-2} \vspace{-5mm}
%
\end{figure*}
The contributions of diagrams Fig. 2 b) and c) are equal to
%
\begin{eqnarray}
%
V^{2}\sum\limits_{\alpha_{1}\alpha_{2}}\sum\limits_{12}\alpha_{1}\alpha_{2}G_{\alpha
\alpha_{1}}^{f(0)}(x|1)G_{\alpha_{1}
\alpha_{2}}^{c(0)}(1|2)G_{\alpha_{2}
\alpha^{\prime}}^{f(0)}(2|x^{\prime}),\nonumber\\
\frac{V^{2}}{2}\sum\limits_{\alpha_{1}\alpha_{2}}\sum\limits_{12}\alpha_{1}\alpha_{2}
G_{\alpha_{1} \alpha_{2}}^{c(0)}(1|2)\left\langle
Tf_{x}^{\alpha}f_{1}^{-\alpha_{1}}f_{2}^{\alpha_{2}}f_{x^{\prime}}^{-\alpha^{\prime}}\right\rangle
_{0}^{ir},\nonumber\label{14}
%
\end{eqnarray}
%
correspondingly.

Between the diagrams for the one-particle $f-$ propagator there
are strong and weak connected ones. The weak connected diagrams
can be separated in two parts by cutting one propagator line. The
sum of all strong connected diagrams for $f-$ electron belong to
the correlation function which is denoted by us as
$\Lambda_{\alpha\alpha^{\prime}}(x|x^{\prime})$ function. The
quantity $\Lambda_{\alpha\alpha^{\prime}}(x|x^{\prime})$ is
defined by the equation
%
\begin{equation}
%
\Lambda_{\alpha\alpha^{\prime}}(x|x^{\prime})=G_{\alpha
\alpha^{\prime}}^{f(0)}(x|x^{\prime})+Z_{\alpha\alpha^{\prime}}(x|x^{\prime}),\label{15}
%
\end{equation}
%
where the function $Z_{\alpha\alpha^{\prime}}(x|x^{\prime})$
contains the contribution of strongly connected diagram based on the
irreducible many-particle Green's functions.

The strong connected part of the $c-$ electron propagator without
the external lines is determined by us as a mass operator for
uncorrelated electrons. This quantity is denoted as
$\Sigma_{\alpha\alpha^{\prime}}(x|x^{\prime})$.

A simple relation exists between these two functions:
%
\begin{equation}
%
\Sigma_{\alpha\alpha^{\prime}}(x|x^{\prime})=V^{2}\alpha\alpha^{\prime}\Lambda_{\alpha\alpha^{\prime}}(x|x^{\prime})\label{16}
%
\end{equation}
%

The analysis of the propagator diagrams permits us to formulate
the following Dyson equation for uncorrelated electron propagator

%
\begin{eqnarray}
%
G_{\alpha \alpha^{\prime}}^{c}(x|x^{\prime})&=&G_{\alpha
\alpha^{\prime}}^{c(0)}(x|x^{\prime})+\\\nonumber
&+&\sum\limits_{\alpha_{1}\alpha_{2}}\sum\limits_{12}G_{\alpha
\alpha_{1}}^{c(0)}(x|1)\Sigma_{\alpha_{1}\alpha_{2}}(1|2)G_{\alpha_{2}
\alpha^{\prime}}^{c}(2|x^{\prime}). \label{17}
%
\end{eqnarray}
%
At the same time we can formulate the Dyson-type equation for
correlated electron propagator $G^{f}$:
%
\begin{eqnarray}
%
G_{\alpha
\alpha^{\prime}}^{f}(x|x^{\prime})=\Lambda_{\alpha\alpha^{\prime}}(x|x^{\prime})+\\\nonumber+
V^{2}\sum\limits_{\alpha_{1}\alpha_{2}}\sum\limits_{12}\alpha_{1}\alpha_{2}\Lambda_{\alpha\alpha_{1}}(x|1)G_{\alpha_{1}
\alpha_{2}}^{c(0)}(1|2)G_{\alpha_{2}
\alpha^{\prime}}^{f}(2|x^{\prime}). \label{18}
%
\end{eqnarray}
%
In equation (15) and (16) as in the previous equations which
contain repeated indices $(1,2)$ is supposed summation by sites
indices $(\mathbf{1},\mathbf{2})$, spin indices
$(\sigma_{1},\sigma_{2})$ and integration by time variables
$(\tau_{1},\tau_{2})$ in the interval $(0,\beta)$.

Unfortunate the Dyson-type equations far correlation function
$\Lambda_{\alpha\alpha^{\prime}}$ and mass operator
$\Sigma_{\alpha\alpha^{\prime}}$ don't exist. Therefore the
calculation of the $c-$ and $f-$ renormalized propagators needs
the approximations based on the summation of special classes of
diagrams.

On the Fig. 3 the skeleton diagrams for the correlation function
$\Lambda_{\alpha\alpha^{\prime}}(x|x^{\prime})$ are depicted .
They demonstrate impossibility to formulate Dyson-type equation
for this function.
\begin{figure*}[t]
%
\centering
\includegraphics[width=0.85\textwidth,clip]{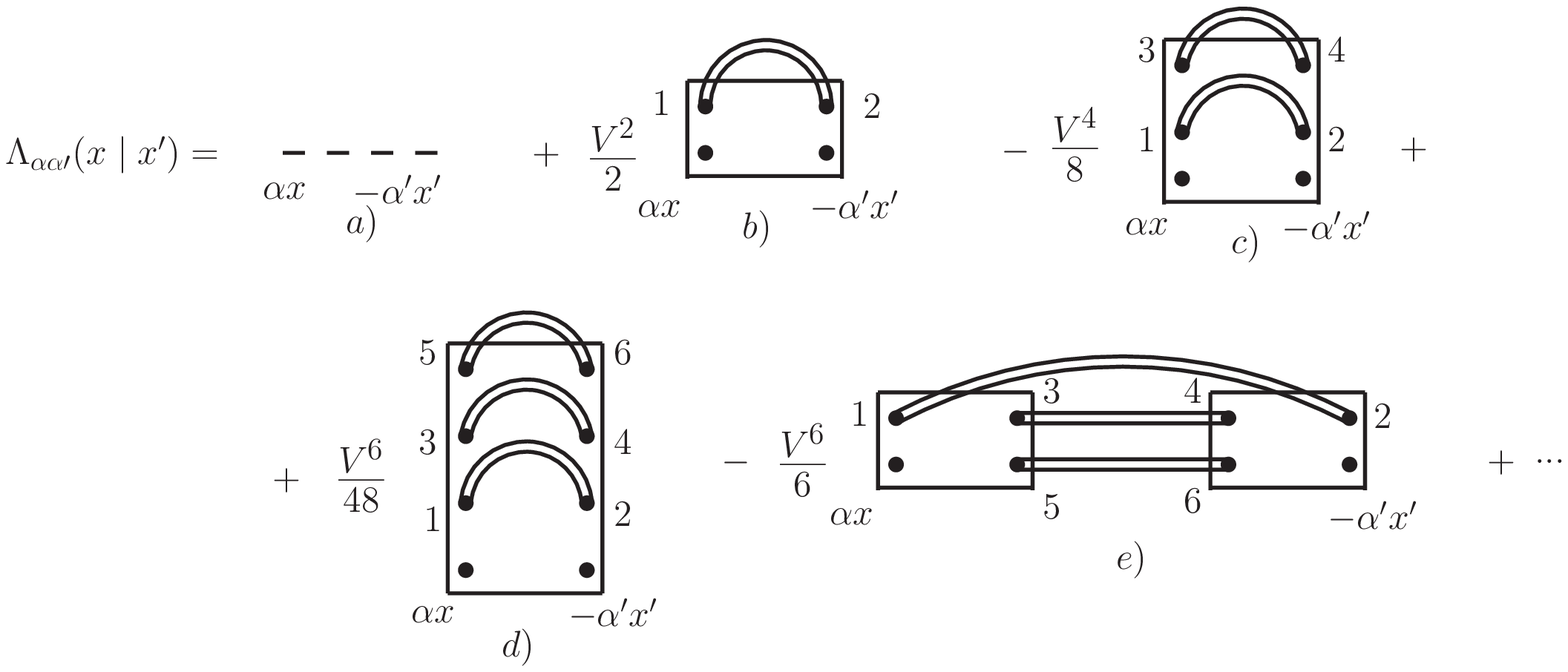}
\vspace{-0mm}
%
\caption{ The skeleton diagrams for correlation function
$\Lambda_{\alpha\alpha^{\prime}}(x|x^{\prime})$. The thin dashed
line is zero-order $f-$ electron Green's function. The rectangles
depict the many-particles irreducible Green's function. The double
solid lines depict the renormalized conduction electron Green's
function $G_{\alpha \alpha^{\prime}}^{c}(x|x^{\prime})$.}
\label{fig-3} \vspace{-5mm}
\end{figure*}
The number of skeleton diagrams depicted on the Fig. 3 for the
correlation function is infinite.

The contribution of the diagrams Fig. 3 b) and c) is the
following:
%
\begin{eqnarray}
%
\frac{1}{2}\sum\limits_{\alpha_{1}\alpha_{2}}\sum\limits_{12}\alpha_{1}\alpha_{2}
G_{\alpha_{1} \alpha_{2}}^{c}(1|2)\left\langle
Tf_{x}^{\alpha}f_{1}^{-\alpha_{1}}f_{2}^{\alpha_{2}}f_{x^{\prime}}^{-\alpha^{\prime}}\right\rangle
_{0}^{ir},\nonumber\\
-\frac{1}{8}\sum\limits_{\alpha_{1}...\alpha_{4}}\sum\limits_{1...4}\alpha_{1}\alpha_{2}\alpha_{3}\alpha_{4}
G_{\alpha_{1} \alpha_{2}}^{c}(1|2)G_{\alpha_{3}
\alpha_{4}}^{c}(3|4)\times\nonumber\\ \times\left\langle
Tf_{x}^{\alpha}f_{1}^{-\alpha_{1}}f_{2}^{\alpha_{2}}f_{3}^{-\alpha_{3}}f_{4}^{\alpha_{4}}f_{x^{\prime}}^{-\alpha^{\prime}}\right\rangle
_{0}^{ir},\nonumber\label{19}
%
\end{eqnarray}
%
correspondingly.

If we take into account only the first term of the right-hand part
of Fig. 3 we obtain the simplest Hubbard I approximation with
consideration only of the chain-type diagrams.

The diagram Fig. 3 b) is the simplest contribution to
correlation function which takes into account the electronic
correlations. The diagrams Fig. 3 b), c) and d) are localized
and their Fourier representations in real space are independent
of momentum. There are also other diagrams of this kind with
irreducible functions $G_{5}^{(0)ir},G_{6}^{(0)ir}$ and so on.
The coefficients before these diagrams are determined by the
number $\frac{1}{2^{n}n!}$, where $2n$ is the perturbation
theory order of diagram. The last diagram of Fig. 3 is not
local and its Fourier representation depends of the momentum.
To dynamical mean field theory only the first group of local
diagrams of Fig. 3 correspond.

The transition of the diagram contribution from superconducting
version to the normal one is realized by the condition of equality
to zero of the sums of all $\alpha-$ indices of every dynamical
quantity. For example such transition of the diagram Fig. 3 b) is
conditioned by the equalities $\alpha_{1}-\alpha_{2}=0$ and
$\alpha-\alpha_{1}+\alpha_{2}-\alpha^{\prime}=0$ with solution
$\alpha_{1}=\alpha_{2}$ and $\alpha=\alpha^{\prime}$.

The summation by $\alpha_{1}$ gives us two equal contributions and
the coefficient before diagram increases twofold and becomes $1$
instead originally $\frac{1}{2}$.

In normal state the correlation function $\Lambda(x|x^{\prime})$
has a form depicted on the Fig. 4.
%
\begin{figure*}[t]
%
\centering
\includegraphics[width=0.85\textwidth,clip]{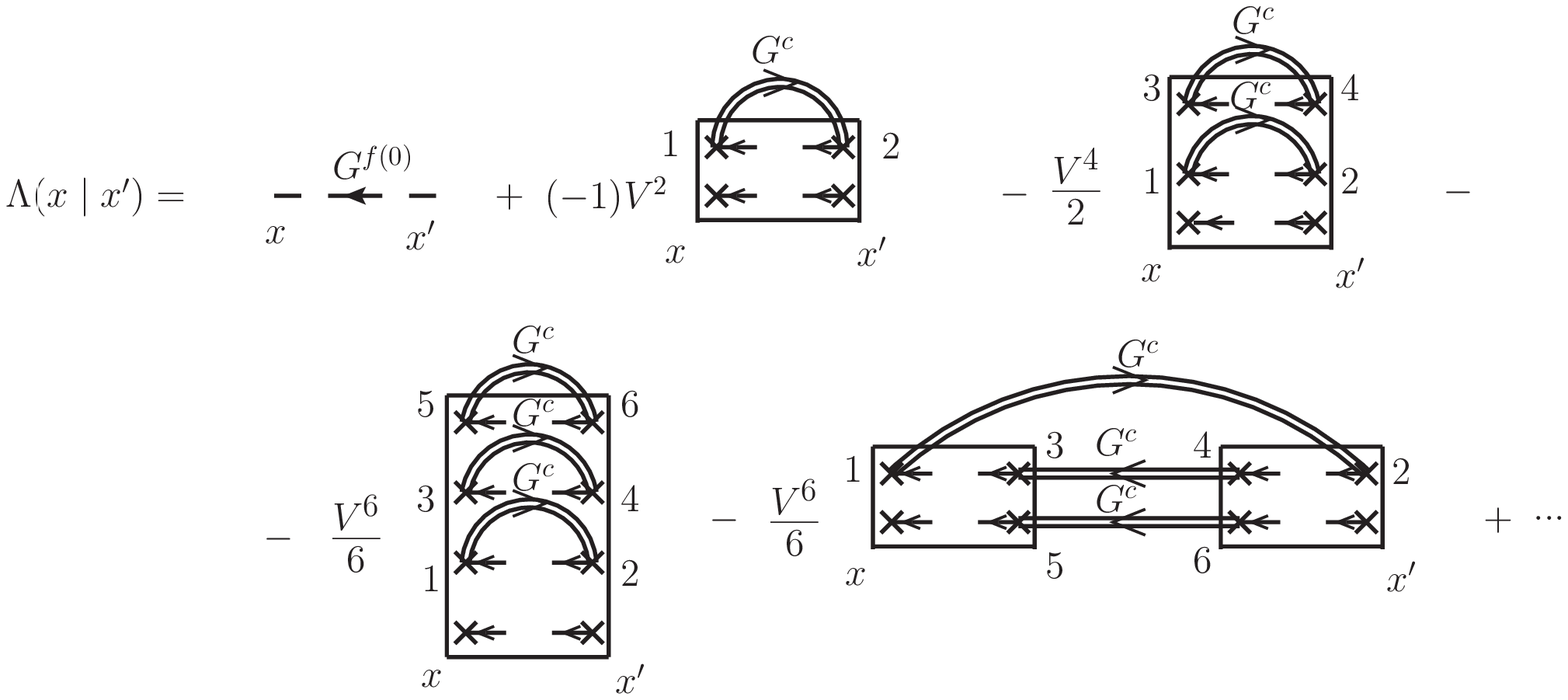}
\vspace{-0mm}
%
\caption{ Correlation function $\Lambda(x|x^{\prime})$ in normal
state. All the lines correspond to normal propagators and have a
direction of propagators and of the arrays in the vertices. }
\label{fig-4} \vspace{-5mm}
\end{figure*}
%

The new coefficient before the diagrams take into account the
existence of different possibilities of transition from
superconducting to normal state.

After discussion of the propagators properties we shall proceed to
the main part of our paper and investigate the properties of
evolution operator average.

\section{Vacuum diagrams}

By using the perturbation theory we have obtained for the
connected part of evolution operator average the contributions
depicted in the Fig. 5.
%
\begin{figure*}[t]
%
\centering
\includegraphics[width=0.85\textwidth,clip]{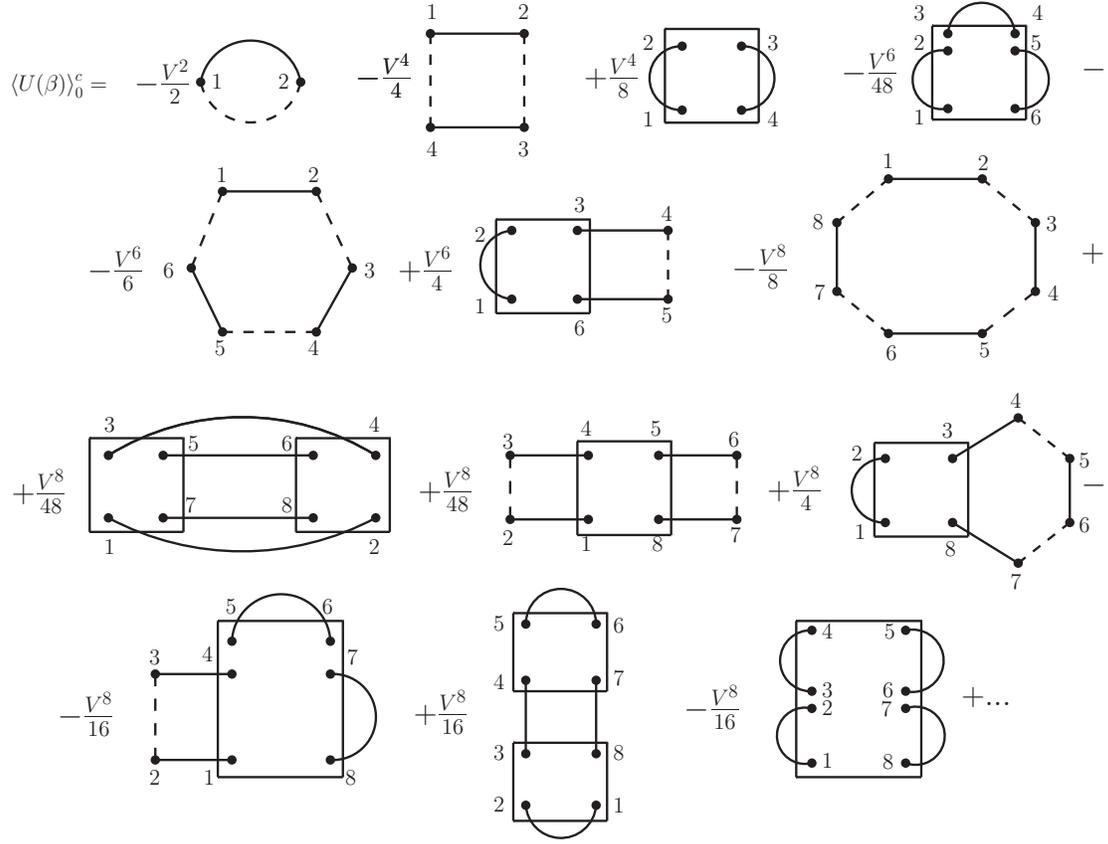}
\vspace{-0mm}
%
\caption{ Vacuum diagrams of first eight orders of perturbation
theory in superconducting state. } \label{fig-5} \vspace{-5mm}
\end{figure*}
%

The contributions of the first, third and fourth diagrams of Fig.
5 are enumerated below:
%
\begin{eqnarray}
%
-\frac{V^{2}}{2}\sum\limits_{\alpha_{1}\alpha_{2}}\sum\limits_{12}\alpha_{1}\alpha_{2}
G_{\alpha_{1} \alpha_{2}}^{c(0)}(1|2)G_{\alpha_{2}
\alpha_{1}}^{f(0)}(2|1),\nonumber\\
\frac{V^{4}}{8}\sum\limits_{\alpha_{1}...\alpha_{4}}\sum\limits_{1...4}\alpha_{1}\alpha_{2}\alpha_{3}\alpha_{4}
G_{\alpha_{1} \alpha_{2}}^{c(0)}(1|2)G_{\alpha_{3}
\alpha_{4}}^{c(0)}(3|4)\times\nonumber\\ \times\left\langle
Tf_{1}^{-\alpha_{1}}f_{2}^{\alpha_{2}}f_{3}^{-\alpha_{3}}f_{4}^{\alpha_{4}}\right\rangle
_{0}^{ir},\nonumber\\\
-\frac{V^{6}}{48}\sum\limits_{\alpha_{1}...\alpha_{6}}\sum\limits_{1...6}\alpha_{1}\alpha_{2}\alpha_{3}\alpha_{4}\alpha_{5}\alpha_{6}
G_{\alpha_{1} \alpha_{2}}^{c(0)}(1|2)G_{\alpha_{3}
\alpha_{4}}^{c(0)}(3|4)\times\nonumber\\\times G_{\alpha_{5}
\alpha_{6}}^{c(0)}(5|6)\left\langle
Tf_{1}^{-\alpha_{1}}f_{2}^{\alpha_{2}}f_{3}^{-\alpha_{3}}f_{4}^{\alpha_{4}}f_{5}^{-\alpha_{5}}f_{6}^{\alpha_{6}}\right\rangle
_{0}^{ir}.\nonumber\label{20}
%
\end{eqnarray}
%

In normal state the diagrams of Fig. 5 are changed. The direction
of propagator lines and of the arrays at vertices points appear
together with new coefficients before the diagrams. This changing
is demonstrated  on the Fig. 6, where some of the diagrams of Fig.
5 are demonstrated.
%
\begin{figure*}[t]
%
\centering
\includegraphics[width=0.85\textwidth,clip]{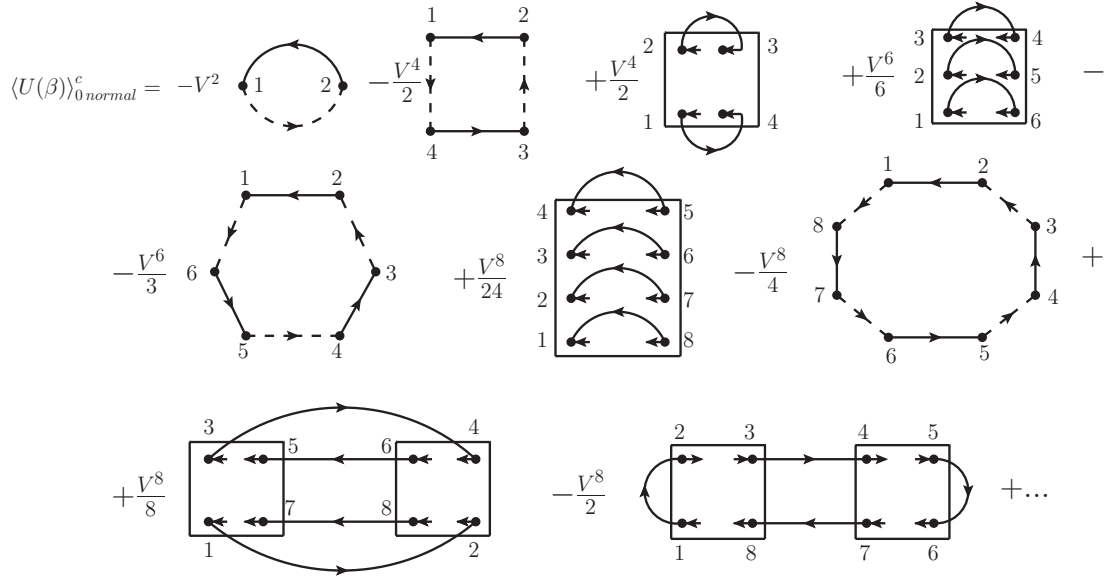}
\vspace{-0mm}
%
\caption{ Some of the vacuum diagrams in normal state of the
system. } \label{fig-6} \vspace{-5mm}
\end{figure*}
Vacuum diagrams in superconducting  and normal states contain the
factor $\frac{1}{n}$, where $n$ is the order of perturbation
theory in which given diagram appears. This factor makes difficult
the investigation of this contributions. In order to remove this
coefficient it is necessary to use the trick of integration by
constant of interaction $V$. The result of such integration is
depicted on the \\Fig. 7.
%
\begin{figure*}[t]
%
\centering
\includegraphics[width=0.85\textwidth,clip]{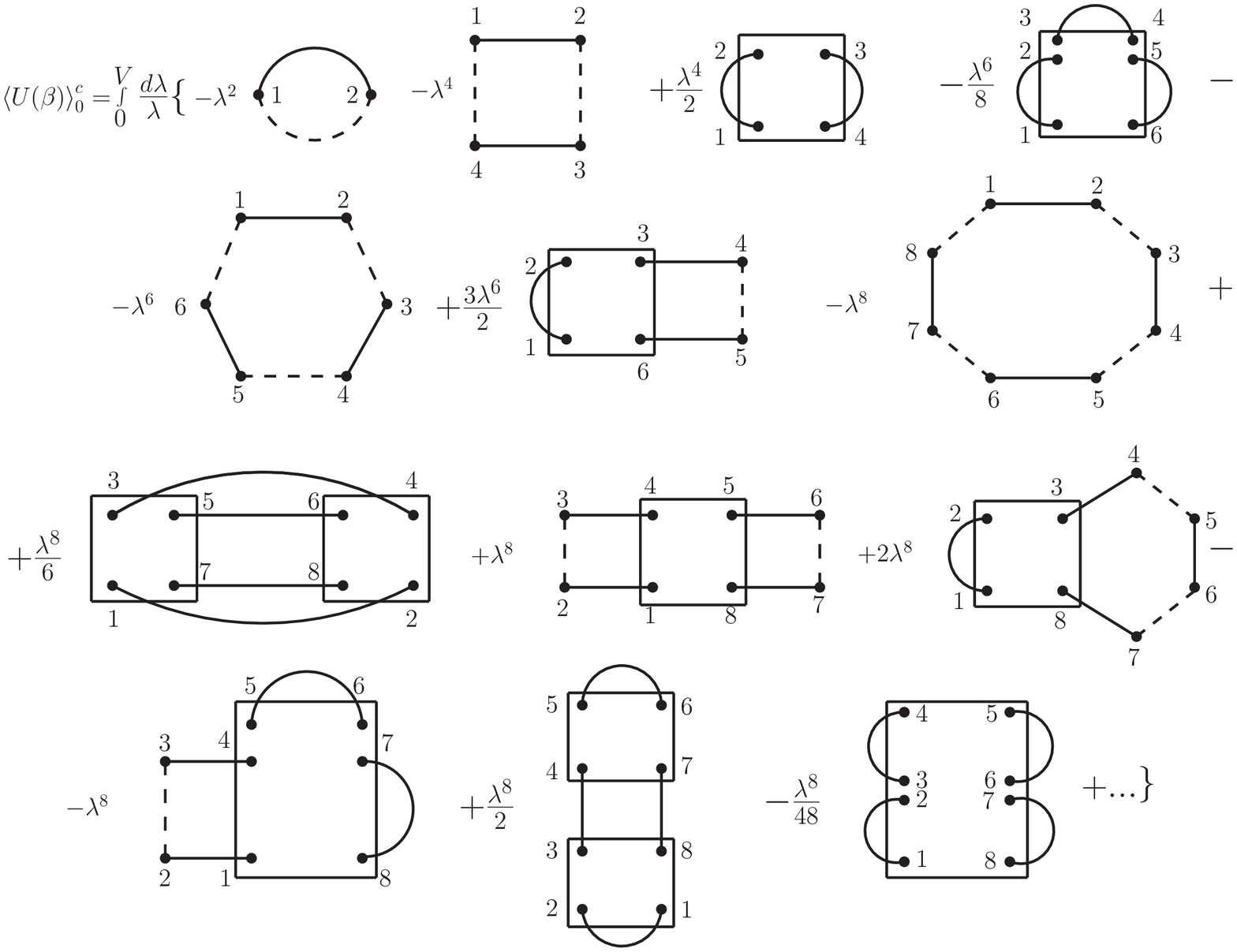}
\vspace{-0mm}
%
\caption{ Vacuum diagrams of the system in superconducting state
after integrating by interaction constant. } \label{fig-7}
\vspace{-5mm}
\end{figure*}
%

On the base of series expansions for renormalized propagators of
the conduction $c-$ electrons (see Fig. 1), localized $f-$
electrons (see Fig. 2) and definition  of the correlation function
$\Lambda_{\alpha\alpha^{\prime}}(x|x^{\prime})$ we can prove that
the contribution of the integrant of Fig. 7 in every order of
perturbation theory can be presented itself  as the product of
some contribution  from $G^{c}$ and some one from $\Lambda$. If
the contribution of $G^{c}$ is of $n_{1}$ order of perturbation
theory and of $\Lambda$ of $n_{2}$ order when the order of
$\left\langle U(\beta)\right\rangle _{0}^{c}$ is equal to $n$ with
the condition $n_{1}+n_{2} +2=n$ which must be satisfied. There
are different possibilities to satisfy this condition and all of
them must be taken into account.

For example there are three possibilities to compose from $G^{c}$
and $\Lambda$ the sixth diagrams of Fig. 7. These possibilities
are enumerated below on the Fig. 8.
%
\begin{figure*}[t]
%
\centering
\includegraphics[width=0.85\textwidth,clip]{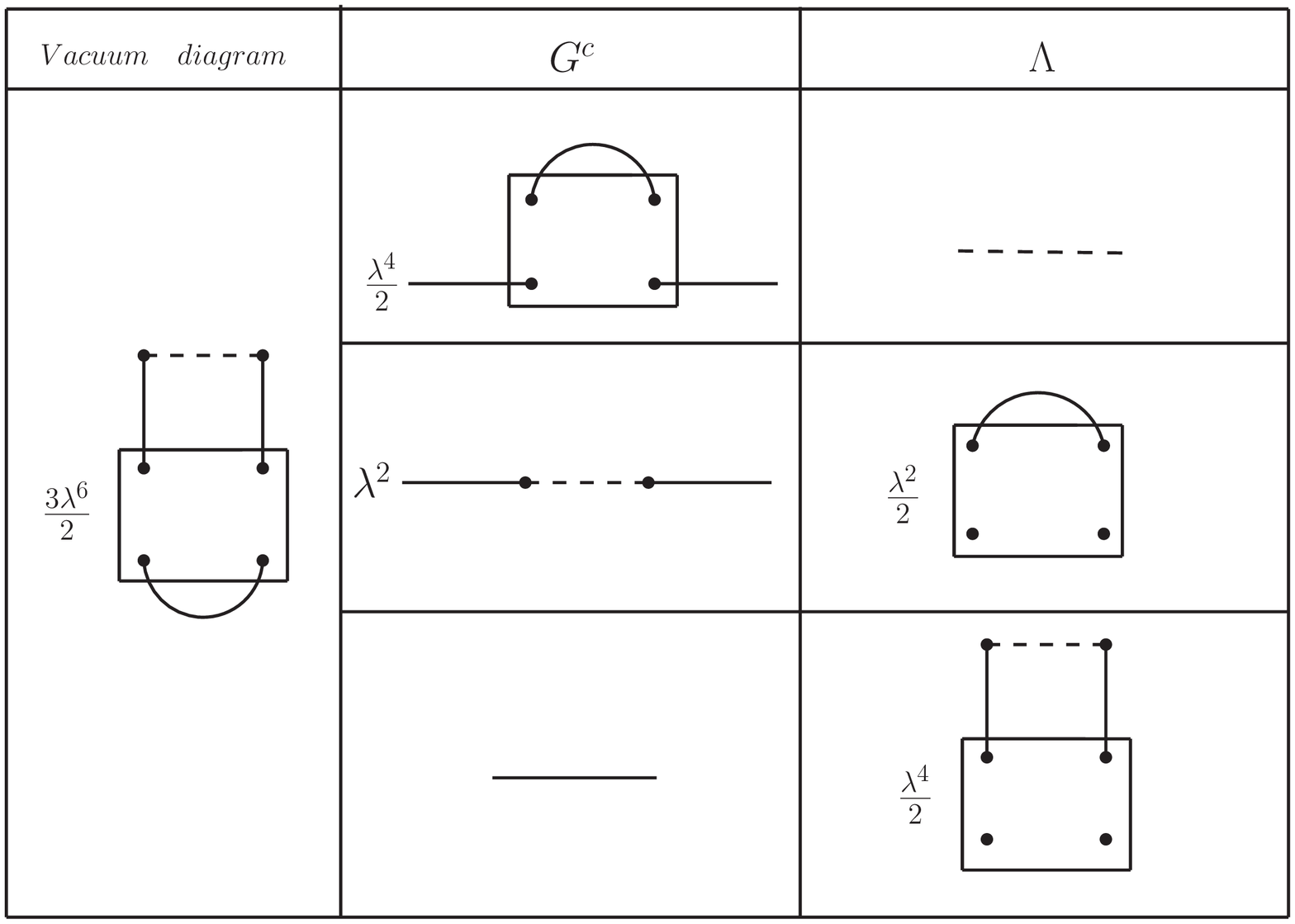}
\vspace{-0mm}
%
\caption{ Three possibilities to organize the vacuum diagram of
sixth order of perturbation theory. The correct coefficient
$\frac{3}{2}$ is obtained by summing all the possibilities. }
\label{fig-8} \vspace{-5mm}
\end{figure*}
%
Other example of vacuum diagram of eighth order of perturbation
theory is presented on the Fig. 9.
%
\begin{figure*}[t]
%
\centering
\includegraphics[width=0.85\textwidth,clip]{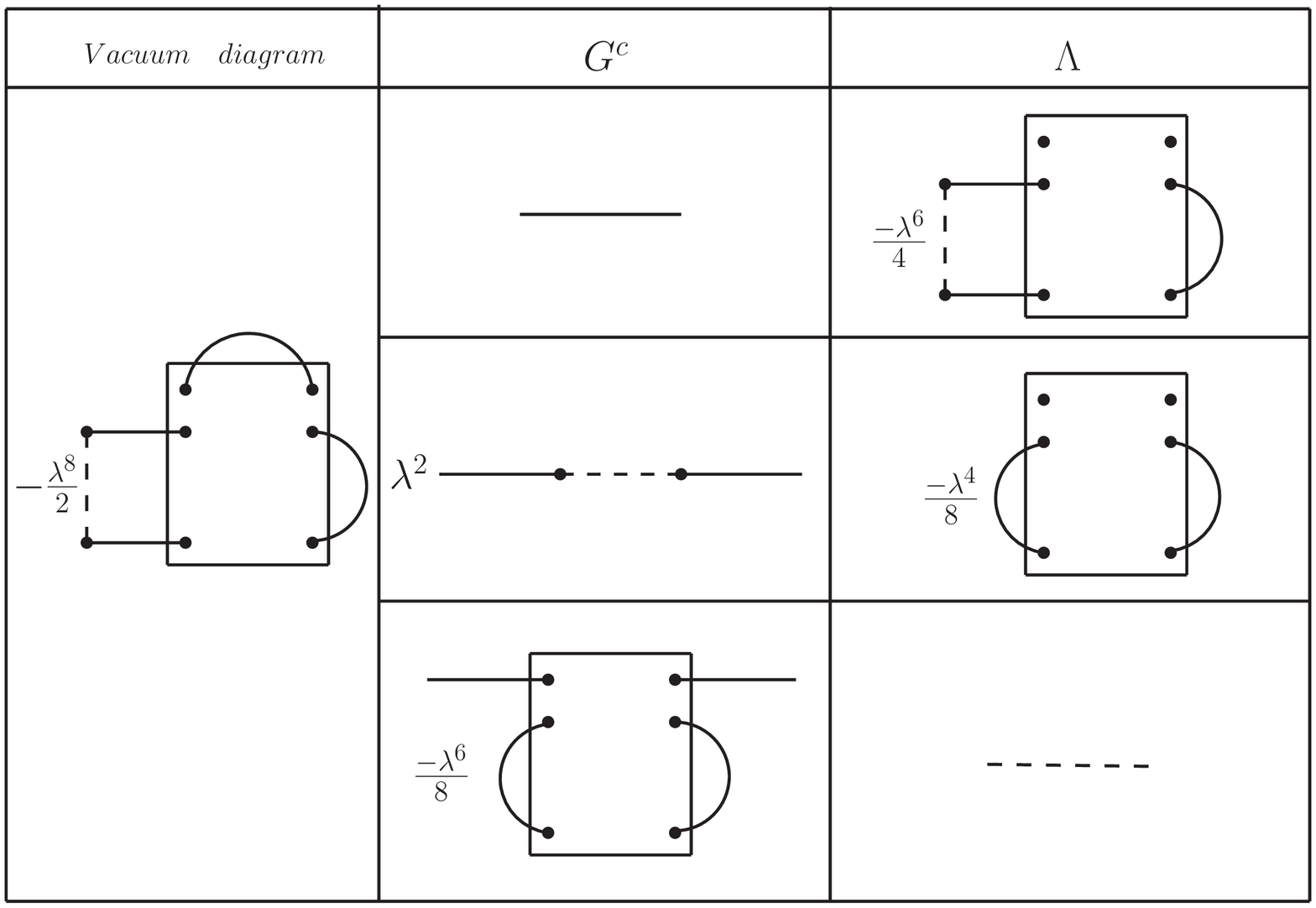}
\vspace{-0mm}
%
\caption{ Three possibilities to obtain one of the vacuum diagram
of eight order of perturbation theory. } \label{fig-9}
\vspace{-5mm}
\end{figure*}
%
Only all the three possibilities give us the correct coefficient
$-\frac{1}{8}-\frac{1}{4}-\frac{1}{8}=-\frac{1}{2}$ of diagram.
Other possibilities don't exist. These examples demonstrate the
general statement that the integrand of the evolution operator
average can be presented itself as a product of
$\lambda^{2}G^{c}\Lambda$ of the form
\begin{widetext}
%
%
\begin{eqnarray}
%
\left\langle U(\beta
)\right\rangle_{0}^{c}=-\int\limits_{0}^{V}\frac{d\lambda
}{\lambda }\sum\limits_{\alpha\alpha^{\prime
}}\sum\limits_{xx^{\prime }}\alpha\alpha^{\prime }\lambda^{2}
G_{\alpha\alpha^{\prime }}^{c}(x|x^{\prime }| \lambda
)\Lambda{\alpha^{\prime }\alpha}
(x^{\prime }|x|\lambda)=\\
=-\int\limits_{0}^{V}\frac{d\lambda }{\lambda
}\sum\limits_{\alpha\alpha^{\prime }}\sum\limits_{xx^{\prime }}
G_{\alpha\alpha^{\prime }}^{c}(x|x^{\prime }|\lambda
)\Sigma{\alpha^{\prime }\alpha} (x^{\prime
}|x|\lambda)=-\int\limits_{0}^{V}\frac{d\lambda }{\lambda }Tr(
\hat{G}^{c}(\lambda )\hat{\Sigma}^{c}(\lambda)). \nonumber
\label{21}
%
\end{eqnarray}
%
\end{widetext}
where the operators $\hat{G}^{c}(\lambda )$ and
$\hat{\Sigma}^{c}(\lambda)$ have the matrix elements
$G_{\alpha\alpha^{\prime }}^{c}(x|x^{\prime }|\lambda )$ and
$\Sigma{\alpha\alpha^{\prime }} (x|x^{\prime }|\lambda)$
correspondingly. Index $\lambda$ underline that these quantities
depend of the auxiliary constant of integration $\lambda$.

Therefore the thermodynamic potential of our system $F$ is equal
to
%
\begin{equation}
%
F=F_{0}+\frac{1}{\beta}\int\limits_{0}^{V}\frac{d\lambda }{\lambda
}Tr[ \hat{G}^{c}(\lambda )\hat{\Sigma}^{c}(\lambda)]. \label{22}
%
\end{equation}
%
This expression for renormalized thermodynamic potential of the
strongly correlated system contains additional integration over
the integration strength $\lambda$ and is awkward because it.
Equation (18) generalizes the result of Luttinger and Ward
$^{[24]}$ proved for non-correlated many-electron system in normal
state.

Our generalization has been obtained for the case of strong
correlations  of special kind which contains one uncorrelated
subsystem and one strongly correlated and we admit also the
existence of superconductivity in both of them.

Luttinger and Ward have proved the possibility to transform this
expression into much more convenient formula without such
integration. For that they used a special functional constricted
from skeleton diagrams the lines of which are the renormalized
electron Green's functions. We shall use the skeleton diagrams of
strongly correlated system which differ essential from Luttinger
and Ward $^{[24]}$ case and transform equation (18) to more
convenient form.

In our strong correlated case we introduce the following
functional
\begin{widetext}
%
%
\begin{eqnarray}
%
Y=-\frac{1}{2\beta }Tr {\{\ln(\hat{G}^{c(0)}\hat{\Sigma} -1)
+\hat{G}^{c}\hat{\Sigma} }\}+ Y ^{\prime }, \label{23}
%
\end{eqnarray}
%
\end{widetext}
which is the generalization of the Luttinger-Ward $^{[24]}$
equation just for the strongly correlated systems. Here operation
$Tr$ use the summation by $\alpha,\sigma,\mathbf{i}$  and
integration by $\tau$.

The quantity $Y ^{\prime }$ contains all peculiarities of the
strongly correlated systems and is presented itself as a sum of
closed linked skeleton diagrams, constructed from irreducible
Green's functions of correlated electrons and full Green's
functions of uncorrelated electrons.

On the Fig. 10 are depicted some of simplest skeleton diagrams for
functional $Y ^{\prime }$. These diagrams depend of the
interaction strength $V$ not only through the factors in the front
of each diagram but also through the dependence of full Green's
function $G^{c}(V)$.
%
\begin{figure*}[t]
%
\centering
\includegraphics[width=0.85\textwidth,clip]{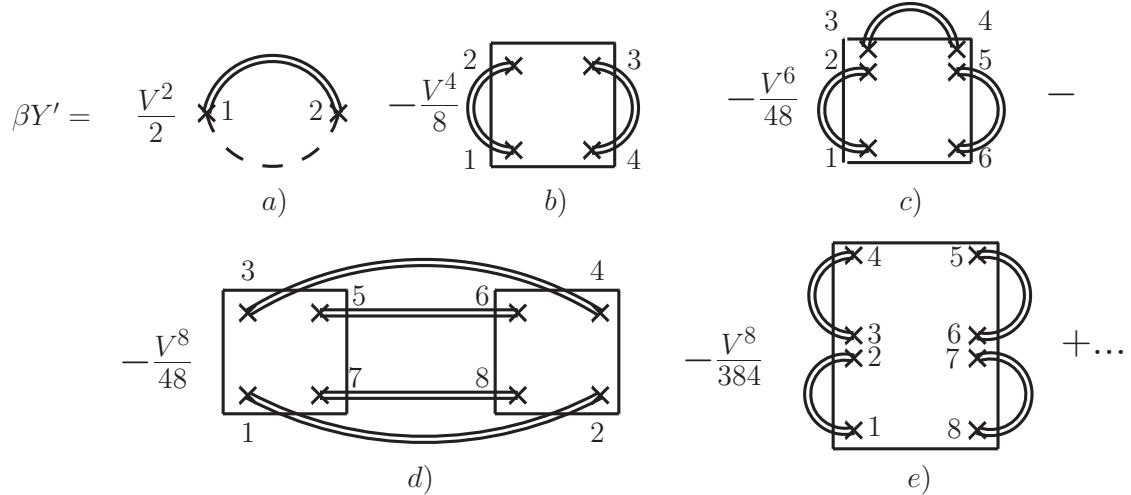}
\vspace{-0mm}
%
\caption{ Skeleton diagrams for functional $Y ^{\prime }$. }
\label{fig-10} \vspace{-5mm}
\end{figure*}
%
The contributions of the diagrams a), b) and c) are following
%
\begin{eqnarray}
%
\frac{V^{2}}{2}\sum\limits_{\alpha_{1}\alpha_{2}}\sum\limits_{12}\alpha_{1}\alpha_{2}
G_{\alpha_{1} \alpha_{2}}^{f(0)}(1,2|V)G_{\alpha_{2}
\alpha_{1}}^{c}(2,1|V),\nonumber\\
-\frac{V^{4}}{8}\sum\limits_{\alpha_{1}...\alpha_{4}}\sum\limits_{1...4}\alpha_{1}\alpha_{2}\alpha_{3}\alpha_{4}
G_{\alpha_{1} \alpha_{2}}^{c}(1,2|V)G_{\alpha_{3}
\alpha_{4}}^{c}(3,4|V)\times\nonumber\\ \times\left\langle
Tf_{1}^{-\alpha_{1}}f_{2}^{\alpha_{2}}f_{3}^{-\alpha_{3}}f_{4}^{\alpha_{4}}\right\rangle
_{0}^{ir},\nonumber\\\
\frac{V^{6}}{48}\sum\limits_{\alpha_{1}...\alpha_{6}}\sum\limits_{1...6}\alpha_{1}\alpha_{2}\alpha_{3}\alpha_{4}\alpha_{5}\alpha_{6}
G_{\alpha_{1} \alpha_{2}}^{c}(1,2|V)G_{\alpha_{3}
\alpha_{4}}^{c}(3,4|V)\times\nonumber\\\times G_{\alpha_{5}
\alpha_{6}}^{c}(5,6|V)\left\langle
Tf_{1}^{-\alpha_{1}}f_{2}^{\alpha_{2}}f_{3}^{-\alpha_{3}}f_{4}^{\alpha_{4}}f_{5}^{-\alpha_{5}}f_{6}^{\alpha_{6}}\right\rangle
_{0}^{ir}.\nonumber\label{24}
%
\end{eqnarray}
%
From Fig. 3 and Fig. 10 it is possible to demonstrate the
following equation:
%
\begin{equation}
%
\frac{\delta \beta Y^{\prime }}{\delta
G_{\alpha_{1}\alpha_{2}}(1,2)}= \frac{\alpha_{1}\alpha_{2}}{2
}V^{2}\Lambda_{\alpha_{2}\alpha_{1}}(2,1)=
\frac{1}{2}\Sigma_{\alpha_{2}\alpha_{1}}(2,1). \label{25}
%
\end{equation}
%
If we take into account only the explicit dependence of the
functional $Y ^{\prime }$ of interaction constant $V$ without
considering the dependence of the full Green's functions
$G^{c}(V)$ from $V$ we shall obtain the other property:
%
\begin{equation}
%
V\beta\frac{dY^{\prime }}{dV}\mid_{G^{c}}=
\sum\limits_{\alpha_{1}\alpha_{2}}\sum\limits_{12}G_{\alpha_{1}
\alpha_{2}}^{c}(2,1|V)\Sigma_{\alpha_{1}\alpha_{2}}(1,2|V)=\nonumber
\\Tr[\hat{\Sigma}\hat{G}^{c}]. \label{26}
%
\end{equation}
%
Because of the necessity to have the functional derivatives over
mass operator $\Sigma$ we shall use the Dyson equation (15)
rewritten in the form
$[x=(\alpha,\mathbf{x},\sigma,\tau)]$:
%
\begin{equation}
%
G^{c(0) -1}(x,x^{\prime })= G^{c
-1}(x,x^{\prime})+\Sigma(x,x^{\prime })\label{27}
%
\end{equation}
%
We obtain
%
\begin{equation}
%
\frac{\delta G^{c}(x,x^{\prime })}{\delta \Sigma(y,y^{\prime
})}=G^{c}(x,y)G^{c}(y^{\prime },x^{\prime }),
 \label{28}
%
\end{equation}
%
and
%
\begin{eqnarray}
%
\frac{\delta}{\delta \Sigma(y,y^{\prime })}
Tr\ln(1-\hat{G}^{c(0)}\hat{\Sigma})= -G^{c}(y^{\prime },y),
\nonumber\\
\frac{\delta}{\delta \Sigma(y,y^{\prime })}
Tr(\hat{G}^{c}\hat{\Sigma})=G^{c}(y^{\prime
},y)+(\hat{G}^{c}\hat{\Sigma}\hat{G}^{c})_{y^{\prime }y}\label{29}
%
\end{eqnarray}
%
By summing these equations we obtain:
%
\begin{eqnarray}
%
\frac{\delta}{\delta \Sigma(y,y^{\prime })}(-\frac{1}{2\beta}) Tr
{\{\ln(\hat{G}^{c(0)}\hat{\Sigma} -1) +\hat{G}^{c}\hat{\Sigma}
}\}=\nonumber \\
=-\frac{1}{2\beta}(\hat{G}^{c}\hat{\Sigma}\hat{G}^{c})_{y^{\prime
}y}. \label{30}
%
\end{eqnarray}
%
On the base of definition of the functional $Y^{\prime}$ Fig. 10
and equation (20) we have
%
\begin{equation}
%
\frac{\delta Y^{\prime }}{\delta \Sigma(y,y^{\prime
})}=\sum\limits_{xx^{\prime }}\frac{\delta Y^{\prime }}{\delta
G^{c}(x,x^{\prime })}\frac{\delta G^{c}(x,x^{\prime })}{\delta
\Sigma(y,y^{\prime
})}=\frac{1}{2\beta}(\hat{G}^{c}\hat{\Sigma}\hat{G}^{c})_{y^{\prime
}y}. \label{31}
%
\end{equation}
%
As a result we obtain the stationarity property of the functional
$Y$:
%
\begin{equation}
%
\frac{\delta Y}{\delta \Sigma(y,y^{\prime })}=0. \label{32}
%
\end{equation}
%

Now we shall discuss the derivative over interaction constant $V$
of functional $Y$. We shall taken into account the stationarity
$Y$ about $\hat{\Sigma}$ and $\hat{G}^{c}$ and equation (21). We
obtain:
%
\begin{eqnarray}
%
V\frac{dY}{dV}=V\frac{\partial Y}{\partial
V}\mid_{\Sigma}+V\frac{\delta Y}{\delta \Sigma}\frac{\partial
\Sigma}{\partial V}=V\frac{\partial Y^{\prime }}{\partial
V}\mid_{\Sigma}=\frac{Tr(\hat{\Sigma}\hat{G}^{c})}{\beta} .
\label{33}
%
\end{eqnarray}
%
From equation (18) we have:
%
\begin{equation}
%
V\frac{dF}{dV}=\frac{Tr(\hat{G}^{c}\hat{\Sigma})}{\beta},
\label{34}
%
\end{equation}
%
and as a consequence we establish
%
\begin{equation}
%
V\frac{dF}{dV}=V\frac{dY}{dV}, \label{35}
%
\end{equation}
%
with the solution
%
\begin{equation}
%
F=Y+\textsc{const}.\nonumber \label{36}
%
\end{equation}
%
This constant is $F_{0}$. Therefore
%
\begin{equation}
%
F=F_{0}+Y,\nonumber \label{37}
%
\end{equation}
%
with the stationary property
%
\begin{equation}
%
\frac{\delta F}{\delta \Sigma(x,x^{\prime })}=0, \label{38}
%
\end{equation}
%
as in superconducting and in normal states.


\section{Heat Capacity [25]}


As a illustration of our results we shall consider the problem of
finding the heat capacity of our strongly correlated system in
normal state and at the low temperatures.

The heat capacity at constant volume $V$ is equal to
%
\begin{equation}
%
C_{V}=T(\frac{\partial S}{\partial T})_{V},\nonumber\label{39}
%
\end{equation}
%
where the entropy $S$ is given by
%
\begin{equation}
%
S=-T(\frac{\partial F}{\partial T})_{\mu,V}.\nonumber \label{40}
%
\end{equation}
%
The quantity $\mu$ is the chemical potential at temperature $T$.

At low temperature we may expand $F(\mu,V,T)$ and $\mu$ in even
powers of the temperature:
%
\begin{eqnarray}
%
F&=&F(\mu_{0},V,0)-\frac{1}{2}\gamma(\mu_{0},V)T^{2}+...,\nonumber\\
\mu(T)&=&\mu_{0}+\mu'T^{2}+..., \nonumber \label{41}
%
\end{eqnarray}
%
where $\mu_{0}$ is the value of chemical potential for correlated
system at $T=0$.

As a result we have
%
\begin{equation}
%
C_{V}=\gamma(\mu_{0},V)T+...\nonumber \label{42}
%
\end{equation}
%
the linear dependence of the heat capacity of the temperature at
low its values. Therefore to evaluate the coefficient
$\gamma(\mu_{0},V)$ it is necessary to obtain the expansion of $F$
in powers of $T$ by using the expression (31), (19) and Fig.10
for $Y'$ obtained by us in previous part of the paper.

We know that the expression (19) for $Y$ and (31) for
thermodynamic potential is stationary with respect to changes in
the proper self-energy $\Sigma_{\mathbf{k}}(i\omega_{0}).$ Therefore because
we are interested in the first corrections to $F$, we can neglect
the explicit temperature dependence of mass operator
$\Sigma(\mathbf{k}|i\omega)$ and $G^{c}(\mathbf{k}|i\omega)$ and replace them by
values $\Sigma(\mathbf{k}|i\omega_{0})|_{T=0}$ and
$G^{c}(\mathbf{k}|i\omega_{0})|_{T=0}$ calculated at $T=0$. Thus the first
correction to the $T=0$ value of $F$ (31) comes only from the
difference between the $\omega_{n}$ sums in expression (31) and
what we would get if we replace them by integrals according to the
equation
%
\begin{equation}
%
\frac{1}{\beta}\sum\limits_{\omega_{n}}=\frac{1}{2\pi}\int\limits_{-\infty}\limits^{\infty}d\omega.\nonumber
\label{43}
%
\end{equation}
%
Now it is necessary to consider the functional $Y'$ on Fig.10.
Since each line of a skeleton diagram of functional $Y'$ contains
an $\omega_{n}$ sum, the total first correction to $Y'$ is
obtained by correcting the computation in each diagram for a
single line and use equation (37) for the other $\omega_{n}$ sums
of the diagram, finally summing over every line.

In such a way we obtain the contribution  of one line of skeleton
diagram multiplied by the number of lines of skeleton diagrams.
This number changes the coefficient before the skeleton diagram of
$Y'$ and new coefficients correspond to new contribution to $Y'$
equal to the self-energy one. This quantity has the form of second
term in right hand part of functional (19) for $Y$ having the
opposite sign. When we combine the both part of functional $Y$
these quantities are reciprocally canceled.

Finally we obtain to the first order for $F$ the equation
\begin{widetext}
%
%
\begin{eqnarray}
%
F&=&F_{0}-\frac{1}{2\beta }\sum\limits_{\sigma
\mathbf{k}}\sum\limits_{\omega_{n}}\ln[G^{c(0)}(i\omega_{n})
\Sigma(\mathbf{k}|i\omega_{n})|_{T=0} -1]e^{i\omega_{n}0^{+}}
=\\\nonumber &=&F_{0}+\frac{1}{2\beta }\sum\limits_{\sigma
\mathbf{k}}\sum\limits_{\omega_{n}}\ln[i\omega_{n}-\epsilon_{\mathbf{k}}]e^{i\omega_{n}0^{+}}-
\frac{1}{2\beta }\sum\limits_{\sigma
\mathbf{k}}\sum\limits_{\omega_{n}}\ln[\Sigma(\mathbf{k}|i\omega_{n})|_{T=0}
-(G^{c(0)}(i\omega_{n}))^{-1}]e^{i\omega_{n}0^{+}}. \label{44}
%
\end{eqnarray}
%
\end{widetext}
We use the Poisson equation for the $\omega_{n}$ sums and write
them as an integral
\begin{widetext}
%
%
\begin{eqnarray}
%
F&=&F_{0}-\frac{1}{2}\sum\limits_{\sigma
\mathbf{k}}\int\limits_{C}\frac{dze^{z0^{+}}}{e^{\beta z}+1} ln[G^{c(0)}(z)
\Sigma(\mathbf{k}|z)|_{T=0} -1],\label{45}
%
\end{eqnarray}
%
\end{widetext}
where $C$ is contour which surrounds in anti clock wise direction
the poles of the function $(e^{\beta z}+1)^{-1}$ in the points
$z=i\omega_{n}=\frac{(2n+1)\pi}{\beta}i$. The term in last
equation proportional to $T^{2}$ is obtained by usual Sommerfeld
technique.

The details of such computation will be discussed in other place.

\section{Conclusions}


We have developed the diagrammatic theory for PAM on the base of
new conceptions proposed by us for strongly correlated electron
systems.

We introduced the notion of correlation function
$\Lambda_{\alpha\alpha^{\prime}}(x|x^{\prime})$  of $f-$
electrons (see Fig. 3) which is the infinite sum of strong
connected irreducible Green's functions and which contains the
most important spin, charge and pairing fluctuations of the
correlated $f-$ electrons. This correlation function determines
the mass operator
$\Sigma_{\alpha\alpha^{\prime}}(x|x^{\prime})$ (14) of the
uncorrelated conduction electrons. The both these quantities
$\Lambda$ and $\Sigma$ permit us to formulate the Dyson
equation for $c-$ electrons (15) and Dyson-type equation (16)
for $f-$ electrons. These results are expressed in general form
appropriate as for normal and as for superconducting state.

We have obtained the skeleton diagrams for $\Lambda$ function and
demonstrated their dependence from irreducible many-particle
Green's functions $G_{n}^{(0)ir}(1,...,2n)$ with all values of $n$
and also of $c-$ electrons full propagators. Thanks the presence
of these irreducible Green's functions it is impossible to
formulate Dyson-type equations for  $\Lambda$ and $\Sigma$
quantities.

The results are appropriate as for normal and as the
superconducting state. Unification of the investigation for the
both phases was possible thanks the introducing of the notion
of quantum charge number $\alpha$ and the rewritten of the
interaction Hamiltonian in such new form.

From Fig. 3 it is clear that the simplest contribution that takes
into account $f-$ electron correlations is reduced to first two
terms of right-hand part of this figure. All the terms of Fig. 3
besides the last one and also other omitted  diagrams like them
are local with Fourier representation independent of momentum.
These terms correspond to the structure of dynamical mean field
theory. Last diagram of Fig. 3 and other more complicated diagrams
with more number of irreducible Green's functions depend of
momentum and take in consideration of the space fluctuations. The
local contributions take into account only of the fluctuations in
time.

We have demonstrated the transition of our diagram from
superconducting to normal state by using the additional conditions
imposed on the charge quantum numbers of which depend the
dynamical quantities.

The special investigation of vacuum diagram has been done after
introducing the auxiliary interaction strength and integration
by it of these diagram contributions. We have proved that this
integrant is equal to the product of two matrices
$\Sigma_{\alpha\alpha^{\prime}}$ and
$G_{\alpha\alpha^{\prime}}^{c}$.

Then we have introduced special functional in the form of skeleton
diagrams and proved it coincidence with thermodynamical potential.
This expression has the property of stationary relative the
changing of the mass operator or full Green's function $G^{c}$ of
conduction electrons.

\begin{acknowledgments}
It is a pleasure acknowledge the discussions with Professor N.M.
Plakida .

\end{acknowledgments}

%

%

\end{document}